# Family of Unconventional Superconductivities in Crystalline Graphene

Junseok Seo[1,†], Armel A. Cotten[2,†], Mingchi Xu[2], Omid Sharifi Sedeh[2], Henok Weldeyesus[2], Tonghang Han[1], Zhengguang Lu[1,3], Zhenghan Wu[1], Shenyong Ye[1], Wei Xu[1], Jixiang Yang[1], Emily Aitken[1], Prayoga P. Liong[1], Zach Hadjri[1], Rasul Gazizulin[4], Kenji Watanabe[5], Takashi Taniguchi[6], Mingda Li[7], Dominik M. Zumbühl[2,*], Long Ju[1,*]

[1]Department of Physics, Massachusetts Institute of Technology, Cambridge, MA 02139, USA
[2]Department of Physics, University of Basel, 4056, Basel, Switzerland
[3]Department of Physics, Florida State University, Tallahassee, FL 32306, USA
[4]Micro-Kelvin Laboratory, University of Florida, Gainesville, FL 32611, USA
[5]Research Center for Electronic and Optical Materials, National Institute for Materials Science, 1-1 Namiki, Tsukuba 305-0044, Japan
[6]Research Center for Materials Nanoarchitectonics, National Institute for Materials Science, 1-1 Namiki, Tsukuba 305-0044, Japan
[7]Department of Nuclear Science and Engineering, Massachusetts Institute of Technology, Cambridge, MA 02139, USA

[†]These authors contributed equally to this work.
*Corresponding author. Email: longju@mit.edu (L.J.), dominik.zumbuhl@unibas.ch (D.M.Z.)

## Abstract

Unconventional superconductors exhibit multiple broken symmetries and exceed the range of the Bardeen-Cooper-Schrieffer (BCS) theory[1]. For instance, time-reversal symmetry can be broken in addition to the gauge symmetry, resulting in superconductors that can be enhanced or induced by a magnetic field[2]. However, such unconventional superconductivities are more vulnerable to impurities than their BCS counterparts[3]—requiring highly ordered and clean material systems to observe them. Crystalline rhombohedral multilayer graphene is a promising platform to explore unconventional superconductivity due to its superior material quality and gate-tunable strong correlation effects[4,5]. Here we report transport measurements of rhombohedral tetralayer and pentalayer graphene, where a spectrum of superconductivities in a clean limit are observed. Three of them (SC2-4) show highly unusual enhancements by magnetic fields: 1. SC2 is strengthened by an in-plane field; 2. SC3 is boosted by a small out-of-plane field; 3. SC4 is induced by an in-plane field. All these superconductors are robust against an in-plane field up to 8.5 Tesla, exceeding the Pauli limit of conventional superconductors[6] by tens of times and suggesting their unconventional nature. Moreover, we observed that proximitized spin-orbit coupling generates a plethora of new superconductors in the phase diagram, while maintaining the high quality of bare rhombohedral graphene. Our work establishes a family of new superconductors in rhombohedral multilayer graphene, which also provides an ideal platform to engineer non-Abelian quasiparticles by proximitizing with quantum anomalous Hall states[7] existing in the same material system[8].

# Main Text

In a strongly correlated system, the normal state of a superconductor is subject to diverse instabilities. This can result in unconventional superconductivity (SC) breaking a symmetry on top of the global gauge symmetry[1] and extending beyond the Bardeen-Cooper-Schrieffer (BCS) theory. Unconventional SCs play a key role in understanding fundamental strongly correlated electron physics and could enable topologically protected qubits for fault-tolerant quantum computation[9]. Despite its importance, it remains one of the biggest open questions in condensed matter physics owing to the intricate interplay between various broken symmetries and their effects on pairing mechanisms—leading to intensive research for decades. In bulk materials[10-17], understanding their doping-dependent behavior relies on comparing different crystals, which can be complicated by varied defect levels and other uncontrolled experimental factors. In contrast, two-dimensional (2D) materials allow for *in situ* tuning of doping by electrostatic gating, which enables comparing effectively hundreds of samples in a reliable fashion. Especially, unconventional SC has been actively explored in moiré superlattices based on graphene and transition metal dichalcogenides (TMDs)[18-25], due to the additional flexibility in structure engineering of interlayer twist angles. Nevertheless, the requirement of a precise twist angle and its inevitable spatial variations[26] in moiré superlattices have posed continued challenges to the device fabrication and reproducing phase diagrams from different devices[27].

Rhombohedral $N$-layer graphene (RNG) has recently emerged as an ideal platform to study unconventional SC. It features simple chemical ingredients and a homogeneous crystalline structure, both facilitating the search of unconventional SCs that are vulnerable to impurities[3]. Furthermore, it hosts a flat low-energy band dispersion ($E \sim k^N$, where $N$ is the number of layers)[28] and large Berry curvature[29], both of which can be further fine-tuned by gate voltages. These features foster enhanced electron correlation effects and various isospin-symmetry-broken states[4,5], which may lead to unconventional SC mediated by mechanisms other than phonons[30-32]. Although SCs in R2G[33-36] and R3G[37-39] have been studied, unconventional SCs in thicker RNG remain less explored. The latter, however, offers several advantages over its thinner counterparts. Firstly, RNGs with $N > 3$ provide richer isospin-symmetry-broken states from which more exotic superconductors can emerge. For example, the layer-antiferromagnet[4,5] and valley-polarized half-metal[40] were not observed in R2G and R3G encapsulated by hexagonal boron nitride (hBN). Secondly, isospin-symmetry-broken parent states can be induced by lower gate electric fields when $N > 3$, reducing the risk of gate-leakage and facilitating further engineering of the devices. Thirdly, RNGs with $N > 3$ have shown a rich family of integer and fractional quantum anomalous Hall (QAH) states[8,41-43]. Should SC be observed in the same RNG, combining them with the observed topological states in a lateral junction could give rise to quasiparticles obeying non-Abelian statistics[7,44,45].

Here, we report electrical transport measurements on hole-doped R4G and R5G, in the simplest possible condition without proximitized SOC and moiré effects. We observed a multitude of superconducting states in both cases, highlighting several unconventional ones in R5G that are enhanced or induced by magnetic fields: 1. SC2 is strengthened by an in-plane magnetic field $B_\parallel$;

2. SC4 is induced by a high $B_\parallel$; 3. SC3 is enhanced by a small out-of-plane magnetic field $B_\perp$ and is robust against $B_\parallel$. These observations are beyond the BCS theory and imply the unconventional nature of these SCs. From a separate R4G/WSe$_2$ device, we observed several new superconducting states induced by SOC effects from a neighboring WSe$_2$ layer. These states are located at relatively low displacement field $D$ and their parent states preserve the long mean-free-path of crystalline graphene—both facts should facilitate unconventional SC through interface engineering.

**SC in rhombohedral multilayer graphene**

Figure 1a and 1b show the $R_{xx}$ maps of R5G and R4G devices at the mixing chamber temperature $T$ of 7 mK, highlighting several regions with vanishing resistance. Here, $R_{xx}$ is defined as the differential four-terminal resistance $dV_{xx}/dI$ at zero DC current. We label the corresponding states as SC1-SC3, as pointed to by the arrows. Figure 1c and 1d show $R_{xx}$ linecuts at fixed $D$ and varied temperatures. $R_{xx}$ decreases to zero as $T$ goes down from ~100 mK in a range of charge densities. The insets show four-terminal differential resistance $dV_{xx}/dI$ as a function of source-drain current $I$ and $B_\perp$. Peaks of $dV_{xx}/dI$ at the critical current are visible at zero $B_\perp$, while both are suppressed at the critical out-of-plane field $B_{c,\perp}$ of ~10 mT. Fermiology analysis based on Shubnikov-de Haas oscillations (see Methods) shows that SC1 is developed at the phase boundary between a full-metal phase with annular Fermi surface and a partially isospin-polarized phase (PIP, Fig. 1e, 1f and Extended Data Fig. 1), which is identical to SC1 in R3G[37]. Temperature-dependent $R_{xx}$, $dV_{xx}/dI$ and fermiology analysis for SC2 in R5G and R4G are presented in Extended Data Fig. 2 and 3.

The sharp drop of $R_{xx}$ by decreased $T$, nonlinear voltage-current relation, and response to $B_\perp$ indicate that SC1 and SC2 in R5G and R4G are superconductors. Their phenomenology is akin to superconductors observed in other 2D materials such as twisted graphene[18-23], twisted WSe$_2$ (refs. [24,25]), and rhombohedral graphene[33-39]. We extract the coherence length $\xi$ of each superconductor as $\xi = \sqrt{\frac{h/(2e)}{2\pi B_{c,\perp}}}$ and the mean-free-path $l$ of its normal state as $l \cong \frac{h}{e^2} \frac{L}{W} \frac{1}{4k_F R_n}$. Here, $R_n$ is normal-state resistance, $k_F \approx \sqrt{\pi n}$ is the Fermi wavevector, $L$ and $W$ are the length and width of the sample, $h$ is Planck's constant and $e$ is the elementary charge[37]. These result in $\xi \approx 200$ nm, $l \approx 1.6$ μm and $\xi \approx 300$ nm, $l \approx 1$ μm for SC1 of R5G and R4G in our devices, respectively, putting these superconductors in the clean limit ($\xi/l \ll 1$). This is in contrast to twisted graphene and WSe$_2$ superconductors, where the normal-state resistance is much higher and the mean-free-path is shorter than the coherence length. Positioned in the clean limit, rhombohedral graphene provides an ideal playground for exploring and studying unconventional SC, which is usually vulnerable to impurities and hard to observe according to Anderson's theorem[3]. We will discuss these unconventional SCs in the following sections.

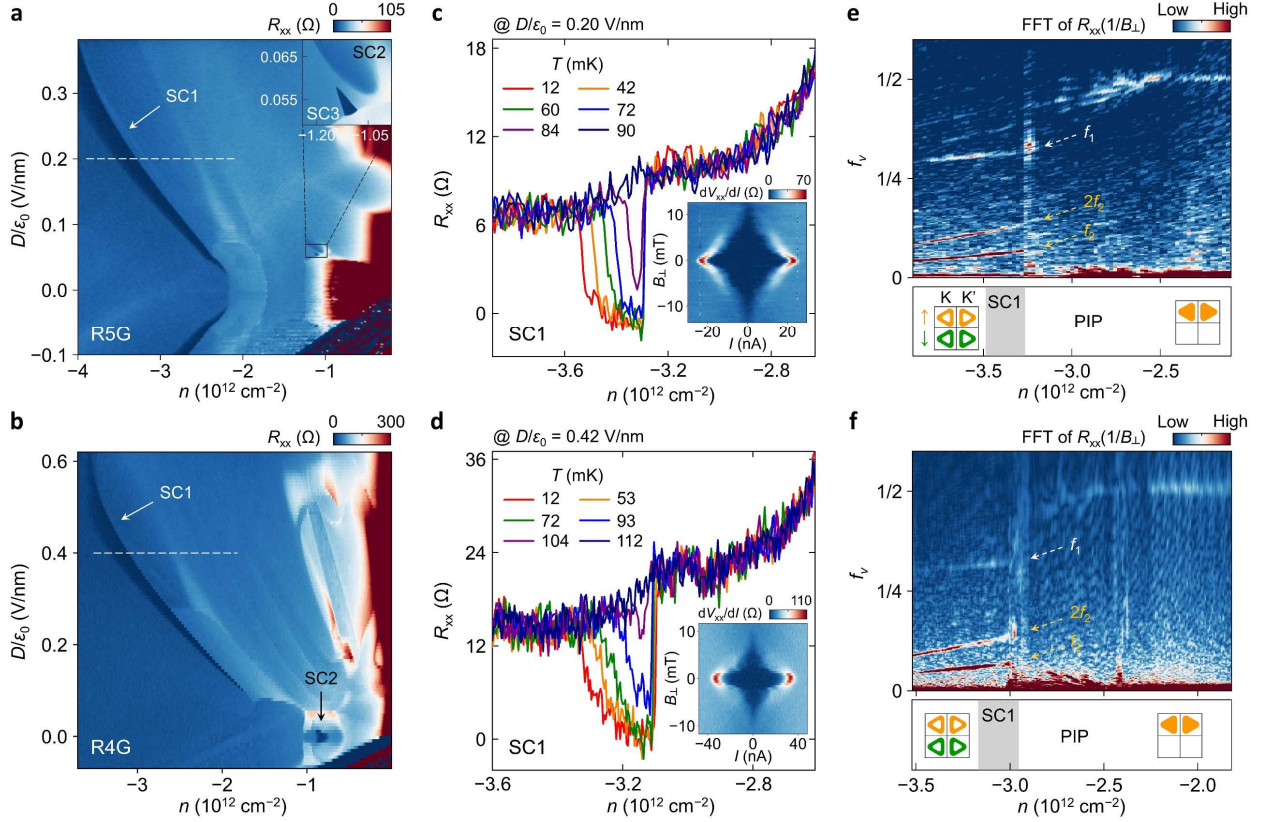

**Figure 1. SC in hole-doped bare RNG. a,b,** Longitudinal resistance $R_{xx}$ as a function of $n$ and $D/\varepsilon_0$ for R5G (**a**) and R4G (**b**), respectively, at zero magnetic field. Inset of **a**: Zoomed-in map corresponding to the black box. Several regions with vanishing $R_{xx}$ are observed, including the SC1 similar to that observed in R3G and three new SC states labeled as SC2 and SC3. **c,d,** $R_{xx}$ as a function of $n$ at varied temperatures and fixed $D/\varepsilon_0$ for SC1 in R5G (**c**) and R4G (**d**), respectively. Inset: Differential resistance $dV_{xx}/dI$ as a function of DC current $I$ and out-of-plane magnetic field $B_\perp$ in R5G ($n = -3.315 \times 10^{12}$ cm$^{-2}$, $D/\varepsilon_0 = 0.20$ V/nm) and R4G ($n = -2.58 \times 10^{12}$ cm$^{-2}$, $D/\varepsilon_0 = 0.28$ V/nm), respectively. Vanishing $R_{xx}$, nonlinear $dV_{xx}/dI$, and the modulation of critical current by $B_\perp$ suggest these are superconducting states. **e,f,** Fourier transform of $R_{xx}(1/B_\perp)$ as a function of $n$ and the normalized frequency $f_v$ (see Methods) along the dashed lines in **a** and **b**, respectively. The relation $f_1 - f_2 = 1/4$ holds for SC1 in both R5G and R4G (gray range in the bottom panel), suggesting an annular-shaped full-metal as the parent state as illustrated. The PIP appears next to this SC1.

## SC2 enhanced and SC4 induced by $B_\parallel$

Having SCs established in R5G and R4G, we further study how they respond to $B_\parallel$ by focusing on R5G. Figure 2a displays $R_{xx}$ maps at $B_\parallel = 0$ and 8 T (see Extended Data Fig. 4 for additional maps) by zooming into the region of SC2 and SC3. We observe two differences comparing the two maps: 1. SC2 is expanded in the $n$-$D$ space at $B_\parallel = 8$ T; 2. a new low-resistance state indicated as SC4 is induced by the high $B_\parallel$. SC4 is well separated from SC2 and SC3 on the map, suggesting it is distinct from the superconductors originally existing at $B_\parallel = 0$ T. We also note that SC3 is robust against the high $B_\parallel$ and eventually connected to SC2 at $B_\parallel = 8.5$ T, as shown in Fig. 2f. SC3 will be discussed in detail in the next section.

Figure 2b shows the differential resistance of SC2 as a function of temperature and current under zero and finite $B_\parallel$ for comparison. Since SC2 shifts to higher $D$ by $B_\parallel$, we take data at the orange and green dot positions in Fig. 2a as they are equally distant from the lower boundary of SC2 in the $n$-$D$ map at the corresponding $B_\parallel$ fields. From $B_\parallel = 0$ to 8 T, the critical current $I_c$ and Berezinskii-Kosterlitz-Thouless (BKT) transition temperature $T_{BKT}$ are increased from $I_c \approx 30$ to $\approx 40$ nA and $T_{BKT} \approx 33$ to $\approx 78$ mK, respectively. Figure 2c shows $R_{xx}$ linecuts along the dashed lines in Fig. 2a and at varied temperatures under $B_\parallel = 0$ and 8 T, respectively. While the dip corresponding to SC2 is fully suppressed at 150 mK for $B_\parallel = 0$ T, it is still observable at the same temperature for $B_\parallel = 8$ T. This confirms the observation of enhanced transition temperature by $B_\parallel$ in Fig. 2b.

The expansion of the phase in the $n$-$D$ space, together with the increase in $T_{BKT}$ and $I_c$, suggests that SC2 is enhanced by $B_\parallel$ and is likely a spin-polarized superconductor. To fully figure out the isospin structure, we first perform Shubnikov-de Haas oscillation measurements and fermiology analyses. Fig. 2d shows the Fourier transform of $R_{xx}(1/B_\perp)$ as a function of $n$ and the normalized frequency $f_v$ (Methods). Two branches with frequencies $f_2$ and $f_3$ obey the relation $f_1 - (f_2 + f_3) \approx 1/2$, and they are merged around the onset of SC2. This suggests the normal state of SC2 is an annular half-metal. The absence of anomalous Hall effect in its normal state, as shown in Fig. 2e, indicates that SC2 is valley-unpolarized. Combining this with the enhancement by $B_\parallel$, we conclude that SC2 arises from a valley-unpolarized and spin-polarized annular half-metal parent state.

In contrast to SC2 which already exists at zero magnetic field, SC4 is only observed at large $B_\parallel$. Fig. 2g shows zoomed-in $R_{xx}$ maps around SC4 at $B_\parallel = 8$ T, featuring a stripe-shaped region with vanishing resistance. Figure 2h compares the differential resistance taken at the orange and black triangles positions in Fig. 2g, corresponding to inside and outside the region of vanishing $R_{xx}$. At the base temperature, differential resistance peaks at threshold currents can only be observed inside SC4. These features disappear as the temperature is increased. Similarly, an out-of-plane magnetic field of around 6 mT also suppresses the nonlinear voltage-current relation of the SC4 state, as shown in Fig. 2i.

The suppression of $R_{xx}$ at low temperatures, nonlinear $dV_{xx}/dI$ and responses to $B_\perp$ for SC4 are consistent with the phenomenology of SC. Interestingly, SC4 is only induced by $B_\parallel$, similar to the SC observed in R2G[33]. We note, however, that the latter is quickly suppressed by a higher $B_\parallel$ (~1 T). In contrast, SC4 shows no sign of being suppressed at up to $B_\parallel = 8.5$ T (Extended Data Fig. 4m), exceeding the Pauli limit[6] of a BCS superconductor with the same transition temperature by a factor of ~65, taking 70 mK as a conservative estimate for $T_{BKT}$. In addition, SC4 in R5G is found at much lower $D$ (~0.15 V/nm versus >0.9 V/nm in R2G), making it easier to access experimentally.

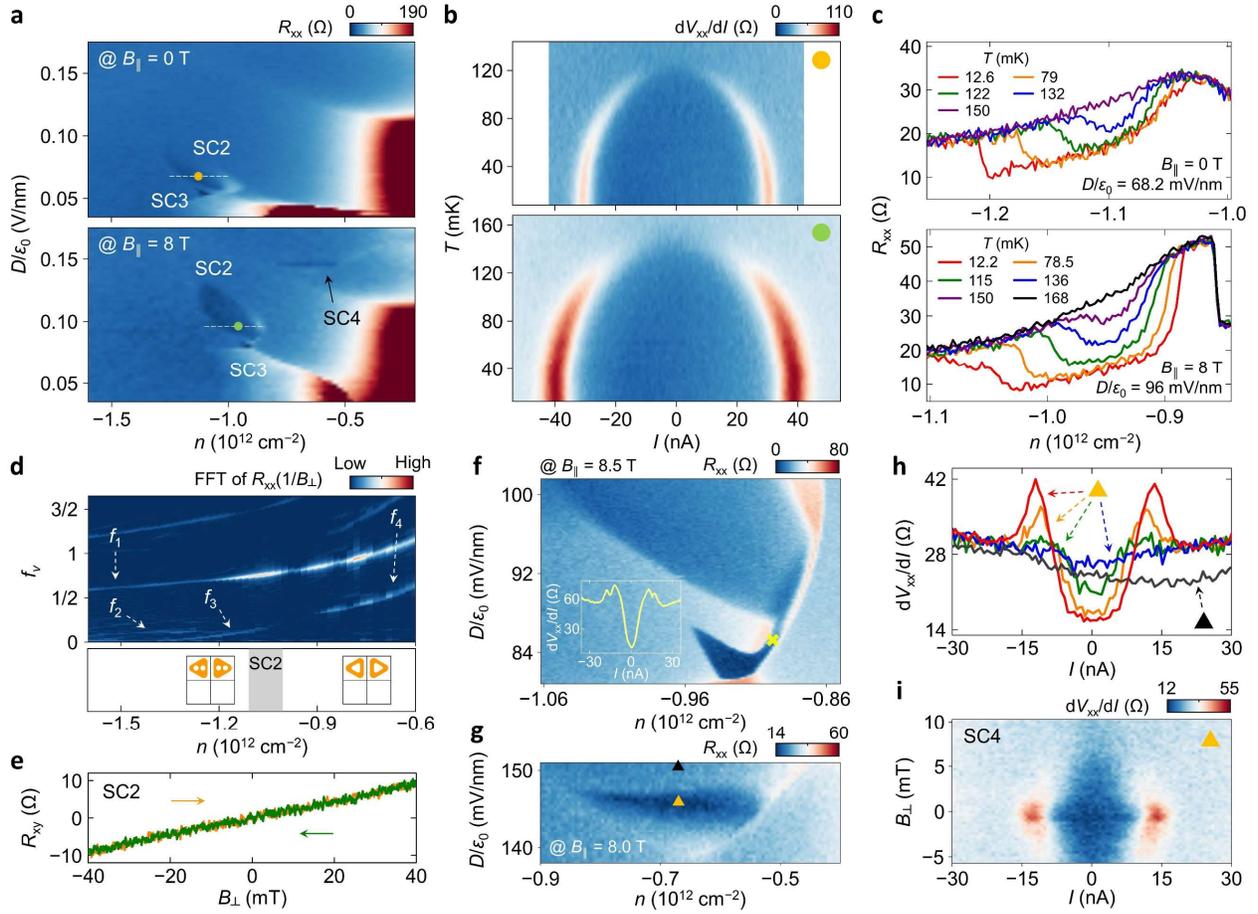

**Figure 2. SC2 and SC4 in R5G, respectively enhanced and induced by $B_\parallel$.** **a**, $R_{xx}$ maps at $B_\parallel = 0$ (top) and 8 (bottom) T. While SC3 is robust, SC2 is enlarged and SC4 is induced by $B_\parallel$. **b**, Differential resistance as a function of $I$ and $T$ at the orange and green dot positions inside the SC2 region in **a**. Both $I_c$ and $T_{BKT}$ are enhanced by $B_\parallel$. **c**, $R_{xx}$ as a function of $n$ at varied temperatures along the white dashed lines in **a**. The $R_{xx}$ dip corresponding to SC2 persists to higher temperatures under finite $B_\parallel$. **d**, Fourier transform of $R_{xx}(1/B_\perp)$ as a function of $n$ and $f_v$. The relation $f_1 - (f_2 + f_3) \approx 1/2$ holds for SC2 (gray range in the bottom panel), suggesting it is born from a half-metal with annular-shaped Fermi surface. **e**, $R_{xy}$ as a function of $B_\perp$ in the forward and backward sweeping directions in the normal state of SC2 ($n = -1.07 \times 10^{12}$ cm$^{-2}$, $D/\varepsilon_0 = 58$ mV/nm). The absence of anomalous Hall effect indicates zero valley polarization. **f**, $R_{xx}$ map at $B_\parallel = 8.5$ T. Inset: Differential resistance as a function of $I$ at the cross point. Nonlinear $dV_{xx}/dI$ between SC2 and SC3 confirms the two superconductors are connected in the $n$-$D$ space. **g**, $R_{xx}$ maps revealing SC4 at $B_\parallel = 8$ T. **h**, Differential resistance as a function of $I$ at $B_\parallel = 8$ T. The red (9 mK), orange (53 mK), green (57 mK) and blue (70 mK) curves correspond to the orange triangle inside the SC4 region. The gray (9 mK) curve corresponds to the black triangle in the neighboring metallic state in **g**. **i**, Differential resistance as a function of $I$ and $B_\perp$ at the orange triangle in **g**, featuring the suppression of $I_c$ by $B_\perp$. Peaks in $dV_{xx}/dI$ at low temperatures and the suppression by small $B_\perp$ are consistent with SC.

## SC3 enhanced by $B_\perp$

As shown in Fig. 2a, SC3 is also robust under a large in-plane magnetic field, similar to SC2 and SC4. Figure 3a shows $R_{xx}$ at $D/\varepsilon_0$ = 52 mV/nm and varied temperatures, featuring nearly zero resistance at the base temperature and sharp rise as $T$ is increased. The inset shows voltage-current relations as $T$ changes, and a characteristic relation as predicted by the BKT theory appears at $T_{BKT}$ = 130 mK. These observations suggest that SC3 is a superconductor.

We perform Shubnikov-de Haas oscillation measurements to understand the Fermi surface structure of the parent state of SC3. Figure 3b shows the Fourier transform of $R_{xx}(1/B_\perp)$ as a function of $n$ and $f_v$. The branch $f_1$ spans across the whole carrier density range, while the low-frequency feature is broken into two segments, $f_2$ and $f_3$. Although the low-frequency branch in the density range of SC3 is unclear, $f_2$ and $f_3$ at its two ends follow relations $f_1 - f_2 \approx 1/2$ and $f_1 - f_3 \approx 1/2$, suggesting a half-metal with annular Fermi surface. Figure 3c shows $R_{xy}$ as $B_\perp$ is swept back and forth for SC3 above $T_c$, showing no anomalous Hall effect. Figure 3d displays a zoomed-in $R_{xx}$ map at $B_\perp$ = 0 mT and $B_\parallel$ = 8.5 T, where SC3 is still observable.

These observations suggest that SC3 is developed from a valley-unpolarized half-metal state with annular Fermi surface. In particular, its behavior under $B_\parallel$ suggests a remarkable Pauli-limit violation ratio (PVR). The maximum $T_{BKT}$ of SC3 is around 130 mK (Fig. 3a), which gives a nominal Pauli limit[6] of $\approx$ 0.24 T. The corresponding PVR for $B_\parallel$ = 8.5 T is >35, which puts SC3 among superconductors with the highest PVR value reported[2,46]. The abnormally high PVR supports the spin-polarized and unconventional nature of SC3.

Figure 3e shows $R_{xx}$ maps at $B_\perp$ = 0 and 1.8 mT. Unlike SC2 and most of the other SCs whose phase space area is shrunk at finite $B_\perp$, SC3 is expanded. Figure 3f shows differential resistance as a function of $I$ and $B_\perp$. The critical current $I_c$ is doubled from $B_\perp$ = 0 to 1.8 mT, before it starts to be suppressed by a higher field. Furthermore, the enhancement of $T_{BKT}$ in the SC3 region from $\approx$ 47 to $\approx$ 76 mK is observed, as shown by the comparison between maps at $B_\perp$ = 0 and 1.8 mT in Fig. 3g. Similar enhancements of $I_c$ and $T_{BKT}$ are observed across the SC3 region, as shown in Extended Data Fig. 5 and 6, and disappear when a finite $B_\parallel$ is applied (Extended Data Fig. 7).

The enhancement of SC by a small $B_\perp$ adds more exoticness to SC3 that is absent from SC2 and SC4. Here we discuss several possible mechanisms for this observation. Firstly, $B_\perp$ can couple to orbital magnetic moments and lower the energy of the SC state versus its competing states. In the parent state of SC3, however, the absence of valley polarization suggests a zero orbital magnetic moment, unlike the chiral SC in electron-doped rhombohedral tetra- and penta-layer graphene[47]. This excludes the orbital magnetism as a possible origin. Secondly, the spin Zeeman energy (~0.2 μeV for $B_\perp$ of 1.8 mT assuming g-factor of 2) can shift the Fermi level to hit a van Hove singularity whose enhanced density of states can lead to stronger SC. However, we did not observe noticeable increase in the $I_c$ of SC3 when similar values of $B_\parallel$ were applied (Extended Data Fig. 8). Thirdly, in a disordered sample exhibiting percolative SC, Josephson junctions formed by superconducting and normal-state islands can lead to the increase in $I_c$ under a tiny $B_\perp$ (ref. [19]). This can occur when the SC is nodal and there is $\pi$ phase difference in the Josephson junction, where this phase shift can be compensated by the $B_\perp$-induced Aharonov-Bohm phase. In this case, the optimal $B_\perp$ value for SC will depend considerably on ($n$, $D$), since the normal-state area connecting two superconducting regions in the Josephson junction will expand as the SC gets weaker. In our experiment, however, the value of $B_\perp$ giving the strongest enhancement of SC3 does not depend on ($n$, $D$) and was comparable between the two devices measured (Extended Data Fig. 5 and 6). We believe further studies are needed to clarify the underlying mechanism of the observed enhancement and how this tiny field substantially increases the critical temperature.

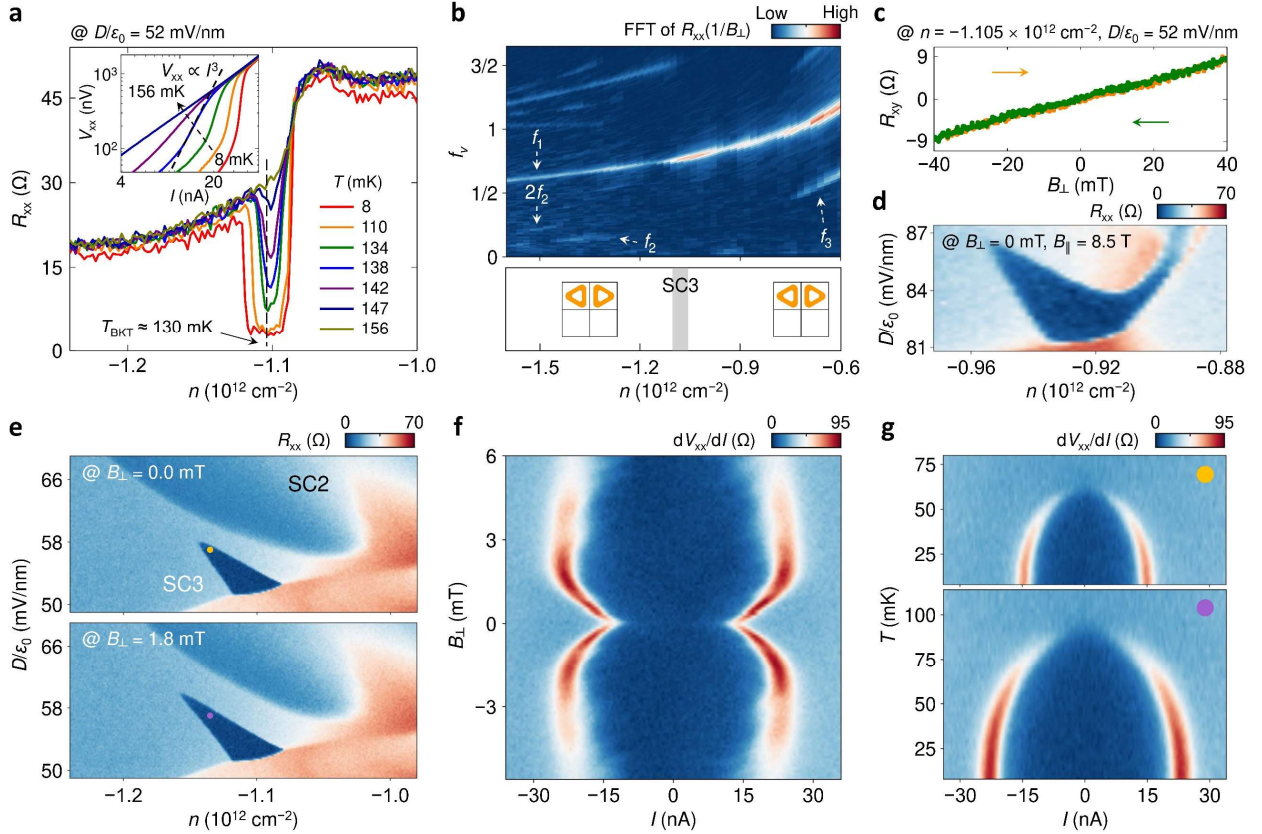

**Figure 3. SC3 in R5G enhanced by $B_\perp$. a**, $R_{xx}$ as a function of $n$ at $D/\varepsilon_0 = 52$ mV/nm and varied temperatures. Inset: Voltage $V_{xx}$ versus current $I$ at $T = 8$, 88, 115, 130, 138 and 156 mK for $n = -1.104 \times 10^{12}$ cm$^{-2}$. The dashed line corresponds to $V_{xx} \propto I^3$ expected from the BKT theory. Vanishing $R_{xx}$ and nonlinear $V_{xx}$-$I$ relation indicate SC3 is a superconductor. **b**, Fourier transform of $R_{xx}(1/B_\perp)$ as a function of $n$ and $f_v$, suggesting SC3 is born from a half-metal with annular-shaped Fermi surface. **c**, $R_{xy}$ as a function of $B_\perp$ in the forward and backward scanning directions in the normal state of SC3. The absence of anomalous Hall effect indicates zero valley polarization. **d**, $R_{xx}$ map at $B_\perp = 0$ mT and $B_\parallel = 8.5$ T. SC3 features an ultrahigh PVR of >35. **e**, $R_{xx}$ maps at $B_\perp = 0.0$ (top) and 1.8 (bottom) mT. The SC3 region is enlarged under finite $B_\perp$. **f**, Differential resistance as a function of $I$ and $B_\perp$ for $n = -1.135 \times 10^{12}$ cm$^{-2}$ and $D/\varepsilon_0 = 57$ mV/nm. $I_c$ is increased by a small $B_\perp$ and suppressed by a larger $B_\perp$. Data in a wider range of $B_\perp$ is shown in Extended Data Fig. 5. **g,** Differential resistance as a function of $I$ and $T$ for $n = -1.135 \times 10^{12}$

cm$^{-2}$ and $D/\varepsilon_0$ = 57 mV/nm under $B_\perp$ = 0.0 (top) and 1.8 (bottom) mT. Both $I_c$ and $T_{BKT}$ are increased by the small $B_\perp$. Data in **e-g** indicate a rare enhancement of SC under a small out-of-plane magnetic field.

**New SCs induced by SOC**

Finally, we examine the impact of proximitized SOC effect on SC in rhombohedral graphene. Figure 4a shows the $R_{xx}$ maps of a separate R4G/WSe$_2$ device at the base temperature. Compared to bare R4G as shown in Fig. 1a, additional states with vanishing $R_{xx}$ are observed and labeled as SC3-7. Fig. 4b to 4e show differential resistance as a function of $I$ and $B_\perp$ for some of these states. In each case, d$V_{xx}$/d$I$ approaches zero at small $I$ and exhibits peaks at critical currents at zero $B_\perp$. These features are suppressed as $B_\perp$ is increased. Extended Data Fig. 9 shows the temperature dependence of $R_{xx}$ and d$V_{xx}$/d$I$ for these states, which are all aligned with the SC phenomenology.

These observations indicate that SC3-7 are superconductors. Phenomenologically, the introduction of SOC from WSe$_2$ proliferates regions with different broken isospin symmetries and renders more SC states at their phase boundaries, akin to the SCs observed in R2G[33-36] and R3G[37-39]. At the same time, SC1 is weakened when the holes are polarized to the WSe$_2$-proximal side of rhombohedral graphene, evidenced by decreases in $T_c$, $I_c$ and $B_{c,\perp}$ (Extended Data Fig. 9 and 10). Together with the observation that SC2 is fully suppressed in R4G/WSe$_2$, the diverse impacts of SOC on SC in R4G are similar to the observations in R3G[39].

We note that the engineering of new SC states by WSe$_2$-proximity keeps the remarkably high-quality of RNG devices. Figure 4f compares normal-state resistance for the SC1 in bare R4G, the WSe$_2$-distant side of R4G/WSe$_2$, and the WSe$_2$-proximal side of R4G/WSe$_2$ at comparable $D$-fields. Although the strength of SC1 varies, all the normal-state square resistance is ~25 Ω/□ for the hole density where SC1 emerges. This suggests that the mean free path, and thus the cleanliness of graphene, is mostly unaffected by the WSe$_2$ layer.

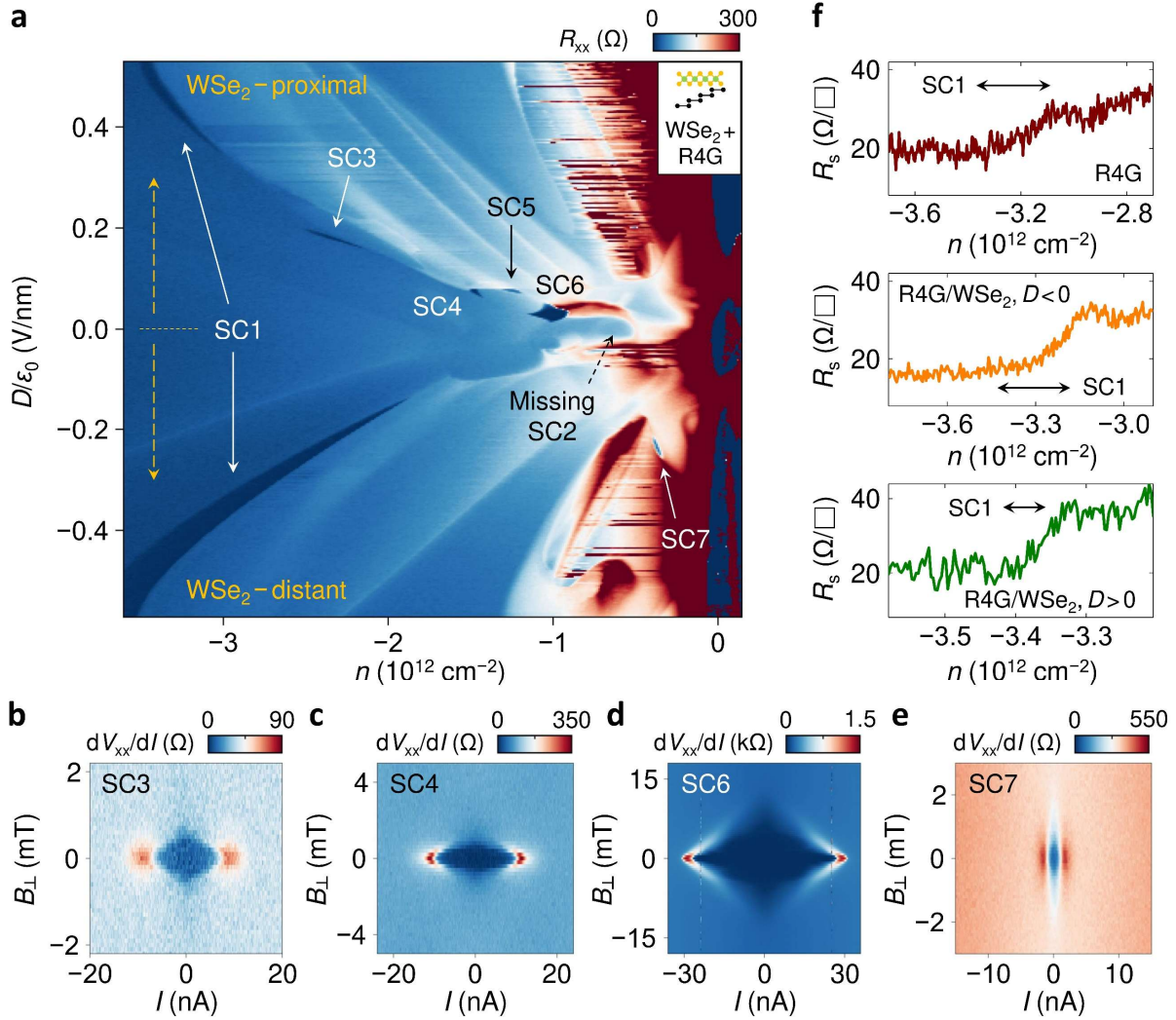

**Figure 4. Proliferation of SC in R4G by proximitized SOC effects. a**, $R_{xx}$ map of a R4G/WSe$_2$ device. Five new superconducting states (SC3-7) emerge on both WSe$_2$-proximal and WSe$_2$-distant sides of displacement field, while the SC2 in Fig. 1a is suppressed. The names of SCs follow the convention of Fig. 1a. **b-e**, Differential resistance as a function of $I$ and $B_\perp$ for SC3 ($n = -2.356 \times 10^{12}$ cm$^{-2}$, $D/\varepsilon_0 = 0.18$ V/nm), SC4 ($n = -1.482 \times 10^{12}$ cm$^{-2}$, $D/\varepsilon_0 = 0.072$ V/nm), SC6 ($n = -1.00 \times 10^{12}$ cm$^{-2}$, $D/\varepsilon_0 = 0.026$ V/nm) and SC7 ($n = -0.38 \times 10^{12}$ cm$^{-2}$, $D/\varepsilon_0 = -0.24$ V/nm). Vanishing $R_{xx}$ (Extended Data Fig. 8) and peaks in d$V_{xx}$/d$I$ and their suppression by $B_\perp$ all suggest the superconducting nature of these states. **f**, Square resistance $R_s$ as a function of $n$ in the normal state of the SC1 of bare R4G (top panel, $D/\varepsilon_0 = 0.42$ V/nm), R4G/WSe$_2$ at negative $D$ (middle panel, $D/\varepsilon_0 = -0.42$ V/nm), and R4G/WSe$_2$ at positive $D$ (bottom panel, $D/\varepsilon_0 = 0.469$ V/nm). Little change in the channel resistance indicates the preserved high quality of graphene with WSe$_2$ at proximity.

## Discussions and outlook

In summary, we have observed a handful of superconducting states in rhombohedral tetra- and penta-layer graphene, featuring three unconventional ones with ultrahigh PVR and unusual responses to magnetic fields. The newly observed SCs show several distinct features, compared to SCs reported in graphene-based materials previously. Firstly, in-plane critical field well exceeding the Pauli limit has been reported in several 2D materials such as (ion-gated) TMDs[46,48-50] and graphene proximitized by TMD[36]. In these materials, strong Ising-type SOC pins the electron spins to the out-of-plane direction and makes the SC state robust against $B_\parallel$. However, bare graphene as used in our work possesses extremely weak SOC[51-53], suggesting spin-polarized SC instead of Ising SC. Secondly, field-enhanced and field-induced SC have been observed in several different materials such as uranium-based compounds[54,55], organic superconductors[56,57] and $Eu_xSn_{1-x}Mo_6S_8$ (ref. [58]). For these materials, either spin fluctuations or the Jaccarino-Peter effect[59] has been suggested to explain the enhancement. In contrast, bare graphene contains neither localized moments nor magnetic atoms, and thus the aforementioned mechanisms cannot be directly applied. Thirdly, the enhancement of SC3 by $B_\perp$ is rarer and more mysterious, considering the absence of valley polarization in its parent state. Further theoretical efforts are needed to explain these field-enhanced and field-induced SCs in R5G.

In addition to the unconventional SCs discussed above, rhombohedral multilayer graphene provides a platform for further engineering the superconductors. One of the promising directions is to proximitize them with integer or fractional QAH states to engineer topological SCs, Majorana fermions and parafermions[7,44,45] using a split-gate device geometry. Significantly, various QAH states with a wide range of Chern numbers and fractional QAH states have been discovered in rhombohedral multilayer graphene neighbored by TMDs[8] or moiré effects from hexagonal boron nitride[41-43]. The co-existence of SCs and such topological states in the same material system can alleviate technical complexities in combining them from different materials. Moreover, many superconductors in our work (including SC2 and SC3 in Fig.1, and SC6 in Fig. 4d) can be reached by a low gate electric field, compared to other superconductors in graphene such as SC in (spin-orbit-coupled) R2G[33-36] and chiral SC in rhombohedral graphene[47,60]. This fact helps accessing the required conditions by mitigating the risk of gate leakage, thereby facilitating to utilize these SCs to realize non-Abelian anyons and fault-tolerant topological quantum computation.

## Methods

**Device fabrications**

Rhombohedral graphene and hBN flakes were prepared onto SiO$_2$/Si substrates by mechanical exfoliation. The rhombohedral domains of graphene were identified and confirmed with an infrared camera[61], near-field infrared nanoscopy[62], and Raman spectroscopy[63], and subsequently isolated by cutting with femtosecond laser. Van der Waals heterostructures were made following a dry transfer procedure. We picked up the top hBN, graphite, middle hBN and graphene using poly(bisphenol A carbonate) film on polydimethylsiloxane, and landed it on a prepared bottom stack consisting of an hBN and graphite bottom gate. The device was then etched into a multi-terminal structure using electron-beam lithography and reactive ion etching. Cr/Au was thermally evaporated for electrical connections to the source, drain, and gate electrodes.

**Electrical transport measurements**

The devices were measured in either a Bluefors LD250 dilution refrigerator at MIT or Leiden MNK126-700 dilution refrigerator at the University of Basel. In the setup at MIT, the DC and AC currents were generated by Keysight 33210A function generator with an AC frequency of 17.77 Hz. Stanford Research Systems SR830 lock-in amplifiers and Basel Precision Instruments (BASPI) SP1004 voltage preamplifiers were used to measure the longitudinal and Hall resistance $R_{xx}$ and

$R_{xy}$. Keithley 2400 source-meters were used to apply top and bottom gate voltages, $V_{TG}$ and $V_{BG}$. In the setup at the University of Basel, MFLI Zurich Instrument lock-in amplifiers were used. The same lock-in amplifier generated the AC current, while the DC current was provided by a BASPI DAC SP927. In addition, the drain-source current was pre-amplified to a voltage signal by SP983c $I$-to-$V$ converter. The applied $V_{TG}$ and $V_{BG}$ were swept to adjust carrier density $n = (C_T V_{TG} + C_B V_{BG})/e$ and displacement field $D = (C_T V_{TG} - C_B V_{BG})/2$, where $C_T$ and $C_B$ are top-gate and bottom-gate capacitance per unit area obtained from the Landau fan diagram.

**Shubnikov-de Haas oscillations and fermiology analyses**

Shubnikov-de Haas oscillations were obtained by measuring $R_{xx}$ as a function of $B_\perp$ at the base temperature. For fermiology analyses, $R_{xx}$ was Fourier-transformed as a function of $1/B_\perp$. The Fourier-transformed data is a function of the frequency $f_{1/B}$. Then, $f_{1/B}$ is normalized by the frequency that corresponds to the full carrier density as $f_\nu = f_{1/B}/(|n|h/e)$.

**Measurements under high $B_\parallel$**

A Cryomagnetics two-axis magnet provided the magnetic fields both in-plane and perpendicular to the sample, with quench boundaries limiting simultaneous fields. The perpendicular component of the in-plane field due to sample misalignment was carefully compensated with the perpendicular magnet using the superconducting sample as an extremely sensitive magnetometer.

Extended Data Fig. 4 shows the evolution of SC2 and SC3 as a function of $B_\parallel$. While the resistance of SC3 is mostly unaffected, $R_{xx}$ for SC2 decreases as $B_\parallel$ up to 5 T is applied. However, the data taken at $B_\parallel = 8.0$ and 8.5 T (Fig. 2f) show higher values of $R_{xx}$ for SC2. To tell if SC2 was indeed weakened, we measured differential resistance at the equivalent ($n$, $D$) point under $B_\parallel = 5$ and 8 T (Extended Data Fig. 4j and 4k). Both the critical current and temperature are identical under these two field values. However, the $dV_{xx}/dI$ curves as functions of $I$ show no change below the mixing chamber temperature $T$ of ~40 mK for $B_\parallel = 8$ T, which is also evident in Fig. 2b. Hence, we conclude that SC2 was not weakened above 5 T but the sample could not be cooled down at 8 T as efficiently as when $B_\parallel = 5$ T and below. One of the possible origins might be vibrations of the sample with reference to the superconducting magnet.

# Acknowledgements

We acknowledge helpful discussions with A. V. Chubukov, C. Yoon and F. Zhang. J.S. was supported by NSF grant DMR-2414725. The device fabrication and transport measurements at MIT were supported by the Nano & Material Technology Development Program through the National Research Foundation of Korea (NRF) funded by the Ministry of Science and ICT (RS-2024-004447252), and the MIT Portugal Program. J.S. acknowledges support from the Jeollanamdo Provincial Scholarship for Study Overseas. T.H. acknowledges a Mathworks Fellowship. The device fabrication for this work was carried out at the Harvard Center for Nanoscale Systems and MIT.nano. Work in Basel was supported by the EU's H2020 Marie Skłodowska-Curie Actions (MSCA) cofund Quantum Science and Technologies at the European Campus (QUSTEC) grant no. 847471, the Swiss National Science Foundation (grant no. 215757), the Georg H. Endress Foundation, the WSS Research Center for Molecular Quantum Systems (molQ) of the Werner Siemens Foundation and the UpQuantVal InterReg. K.W. and T.T. acknowledge support from the JSPS KAKENHI (Grant numbers 21H05233 and 23H02052), the CREST (JPMJCR24A5), JST and World Premier International Research Center Initiative (WPI), MEXT, Japan.

**Author contributions:** L.J. and D.M.Z. supervised the project. T.H., Z.W., S.Y., W.X., E.A. and P.P.L. fabricated the devices. J.S. performed measurements at MIT with assistance from T.H., Z.L. and J.Y.; J.S. carried out measurements at the University of Basel with assistance from A.A.C., M.X., O.S.S., H.W., Z.H. and D.M.Z.; J.S. and L.J. wrote the paper with input from all authors.

**Competing interests:** D.M.Z. is a co-founder of Basel Precision Instruments. The other authors declare no competing financial interests.

**Data availability:** The data that support the findings of this study are available from the corresponding author on reasonable request.

# Extended Data Figures

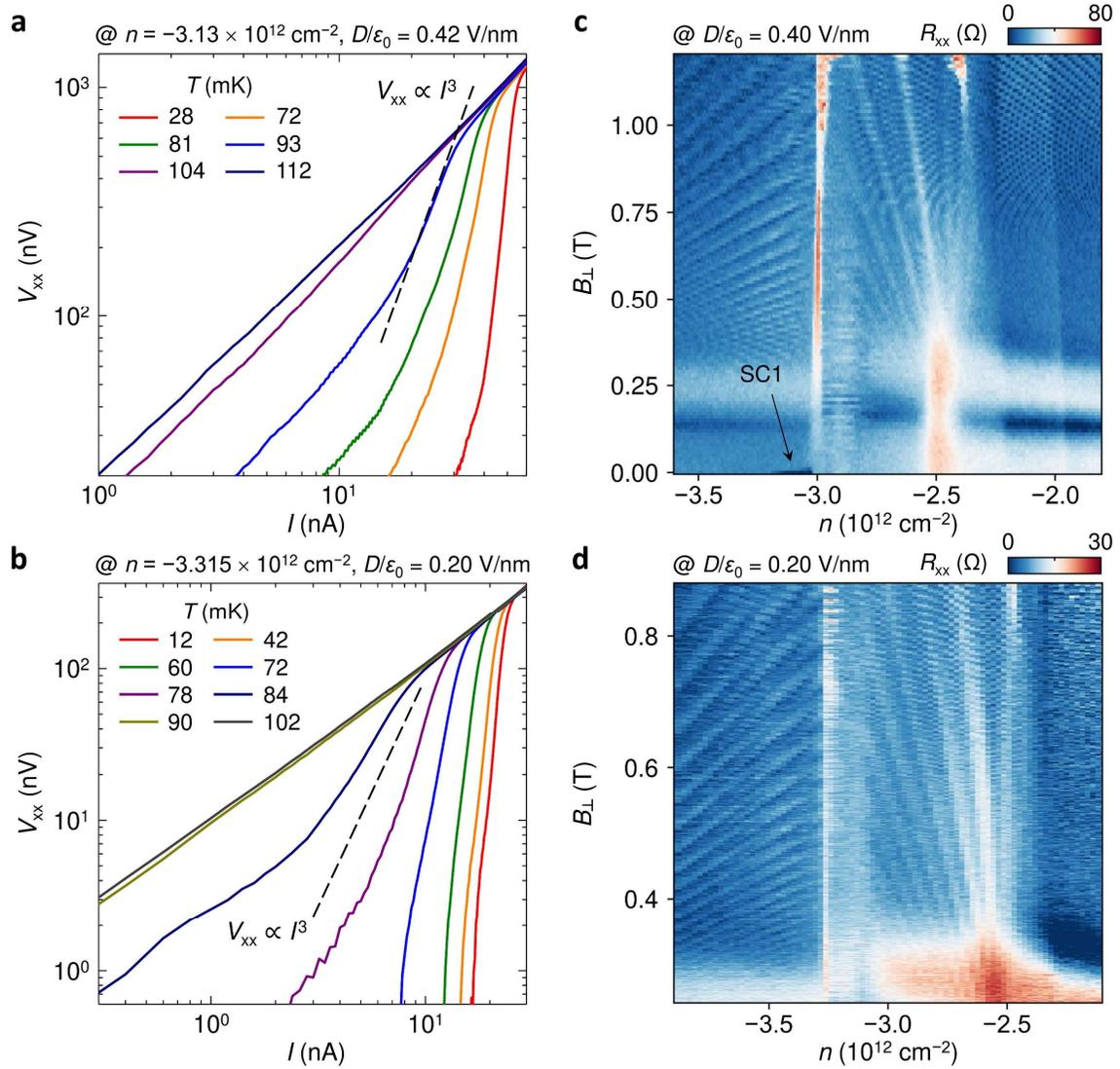

**Extended Data Figure 1. Additional data for SC1 in R4G and R5G. a,b,** Voltage $V_{xx}$ versus current $I$ at varied temperatures for the SC1 of (**a**) R4G and (**b**) R5G. **c,d,** $R_{xx}$ as a function of $n$ and $B_\perp$ for (**c**) R4G and (**d**) R5G. The data in **c** and **d** have been used for the Fourier transform in Fig. 1e and 1f.

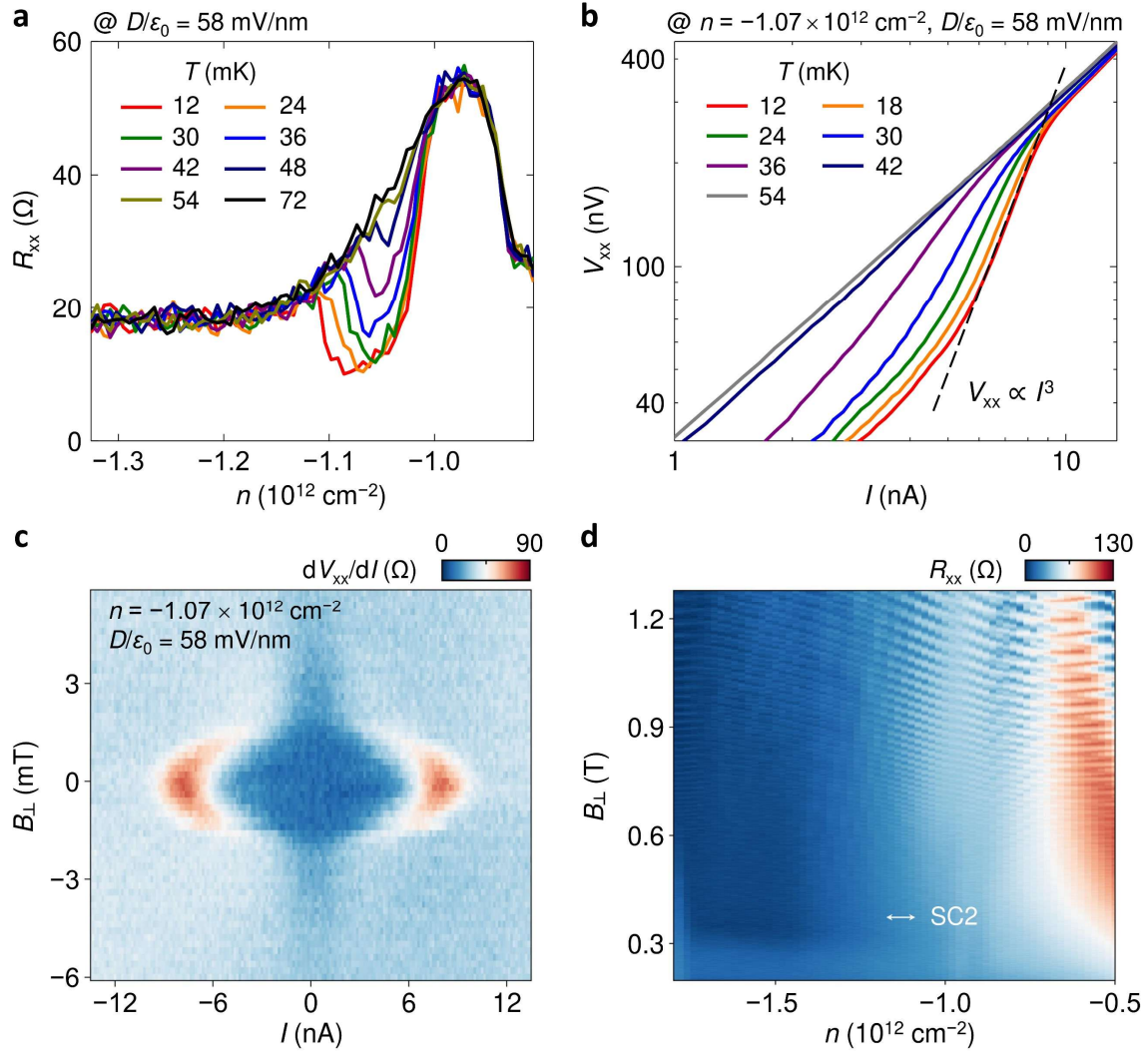

**Extended Data Figure 2. SC2 in R5G. a,** $R_{xx}$ as a function of $n$ at varied temperatures. **b,** Voltage $V_{xx}$ versus current $I$ at varied temperatures. The dashed line denotes when $V_{xx}$ is proportional to $I^3$, which corresponds to $T_{BKT} \approx 12$ mK. **c,** Differential resistance as a function of $I$ and $B_\perp$ at $n = -1.07 \times 10^{12}$ cm$^{-2}$ and $D/\varepsilon_0 = 58$ mV/nm. The suppression of $R_{xx}$, nonlinear $dV_{xx}/dI$, and its disappearance by $B_\perp$ suggest SC2 in R4G is a superconductor. In addition, $B_{c,\perp} \approx 1.5$ mT and $R_n \approx 30$ Ω at this ($n$, $D$) point, which puts the SC2 around the clean limit ($\xi \approx 470$ nm and $l \approx 1.1$ μm). **d,** Landau fan diagrams taken at $D/\varepsilon_0 = 63$ mV/nm as a function of $n$. This data has been used for the Fourier transform in Fig. 2d.

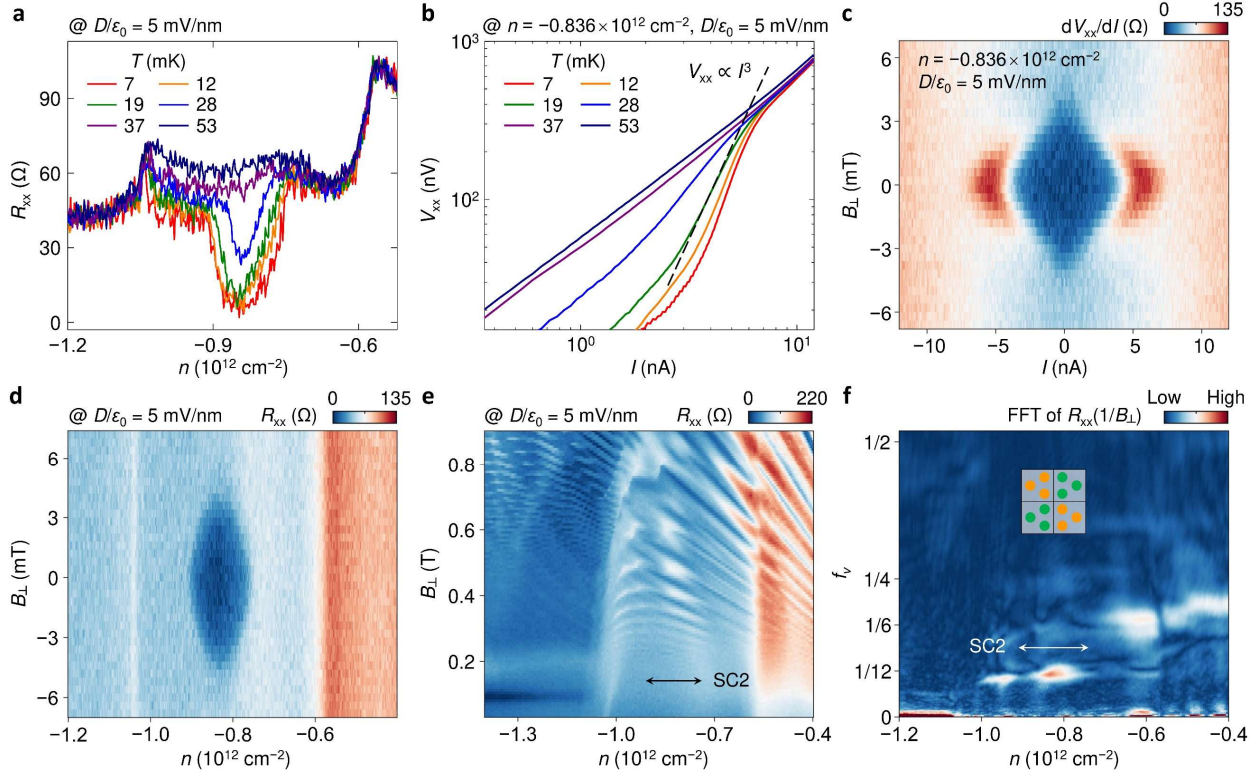

**Extended Data Figure 3. SC2 in R4G. a,** $R_{xx}$ as a function of $n$ at varied temperatures. **b,** Voltage $V_{xx}$ versus current $I$ at varied temperatures. The dashed line denotes when $V_{xx}$ is proportional to $I^3$, which corresponds to the BKT transition temperature $T_{BKT} \approx 19$ mK. **c,** Differential resistance as a function of $I$ and $B_\perp$ at $n = -0.836 \times 10^{12}$ cm$^{-2}$ and $D/\varepsilon_0 = 5$ mV/nm. Vanishing $R_{xx}$, nonlinear $dV_{xx}/dI$, and their quick suppression by $B_\perp$ suggest SC2 in R4G is a superconductor. In addition, $B_{c,\perp} \approx 3$ mT and $R_n \approx 60$ Ω at this ($n$, $D$) point, which puts the SC2 near the clean limit ($\xi \approx 330$ nm and $l \approx 660$ nm). **d,** $R_{xx}$ as a function of $n$ and $B_\perp$ at $D/\varepsilon_0 = 5$ mV/nm. **e,** Landau fan diagrams taken at $D/\varepsilon_0 = 5$ mV/nm as a function of $n$. **f,** Fourier transform of $R_{xx}(1/B_\perp)$ as a function of $n$ and $f_v$. There is a peak at $f_v = 1/12$, which suggests the normal state of SC2 in R4G is a trigonally warped full metal.

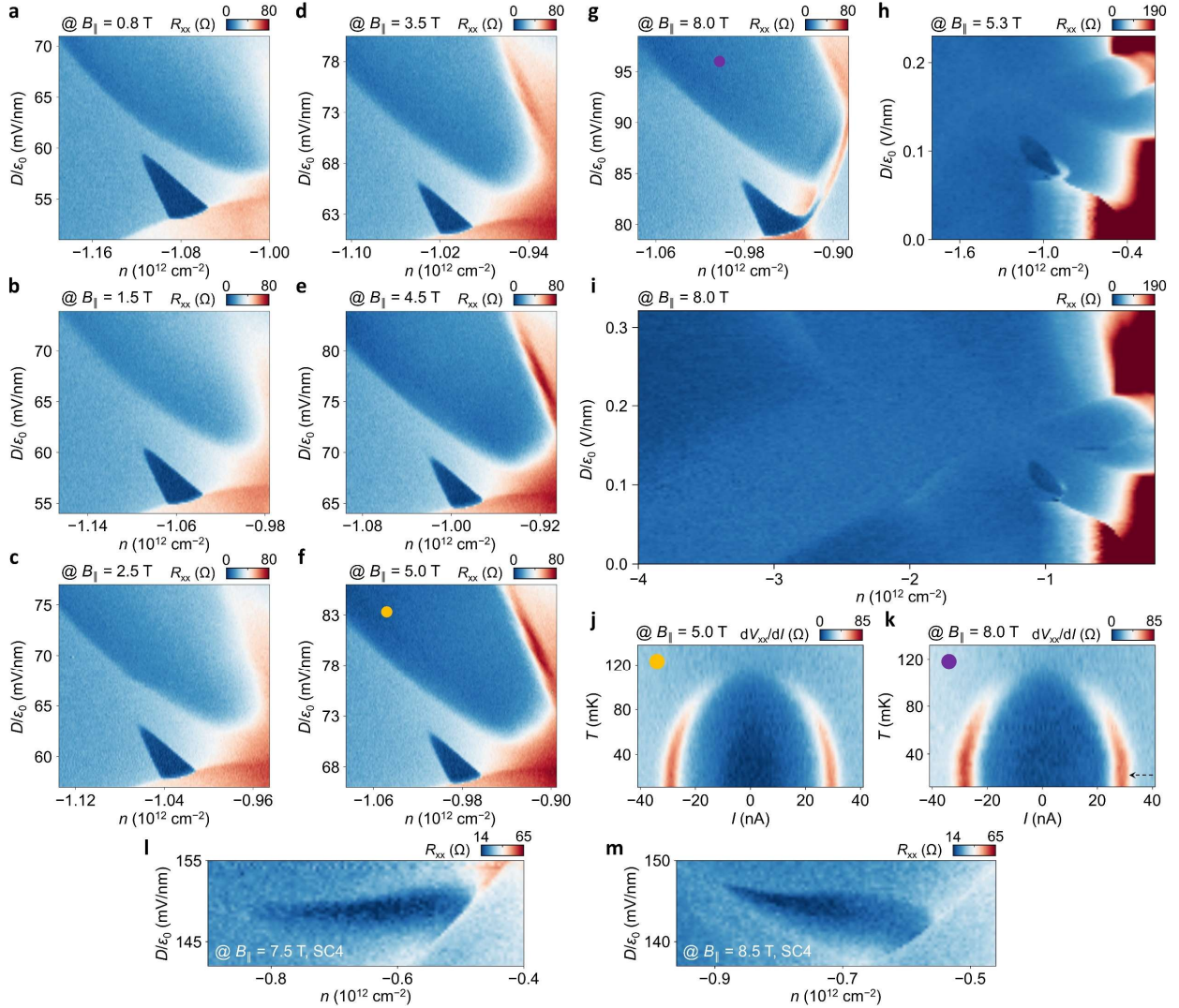

**Extended Data Figure 4. SC in R5G under $B_\parallel$. a-g,** $R_{xx}$ maps as a function of $n$ and $D/\varepsilon_0$ focusing on SC2 and SC3 at $B_\parallel = 0.8$ T (**a**), $B_\parallel = 1.5$ T (**b**), $B_\parallel = 2.5$ T (**c**), $B_\parallel = 3.5$ T (**d**), $B_\parallel = 4.5$ T (**e**), $B_\parallel = 5.0$ T (**f**), and $B_\parallel = 8.0$ T (**g**). **h,** $R_{xx}$ map at $B_\parallel = 5.3$ T. SC4 is not yet developed. **i,** $R_{xx}$ map in a large $n$-$D/\varepsilon_0$ space at $B_\parallel = 8.0$ T. SC1 is fully killed at this magnetic field. **j,k,** Differential resistance as a function of $I$ and $T$ at the equivalent $(n, D)$ point under $B_\parallel = 5$ T (**j**, $n = -1.048 \times 10^{12}$ cm$^{-2}$ and $D/\varepsilon_0 = 83.3$ mV/nm, yellow dot in **f**) and $B_\parallel = 8$ T (**k**, $n = -1.001 \times 10^{12}$ cm$^{-2}$ and $D/\varepsilon_0 = 96$ mV/nm, purple dot in **g**). Both $I_c$ and $T_c$ are not altered, but the $dV_{xx}/dI$ curve does not change below ~40 mK for $B_\parallel = 8$ T (black arrow). This signals the difficulty in cooling down the sample. **l,m,** $R_{xx}$ maps focusing on SC4 at $B_\parallel = 7.5$ T (**l**) and $B_\parallel = 8.5$ T (**m**). SC4 was not suppressed up to the highest field available in the magnet.

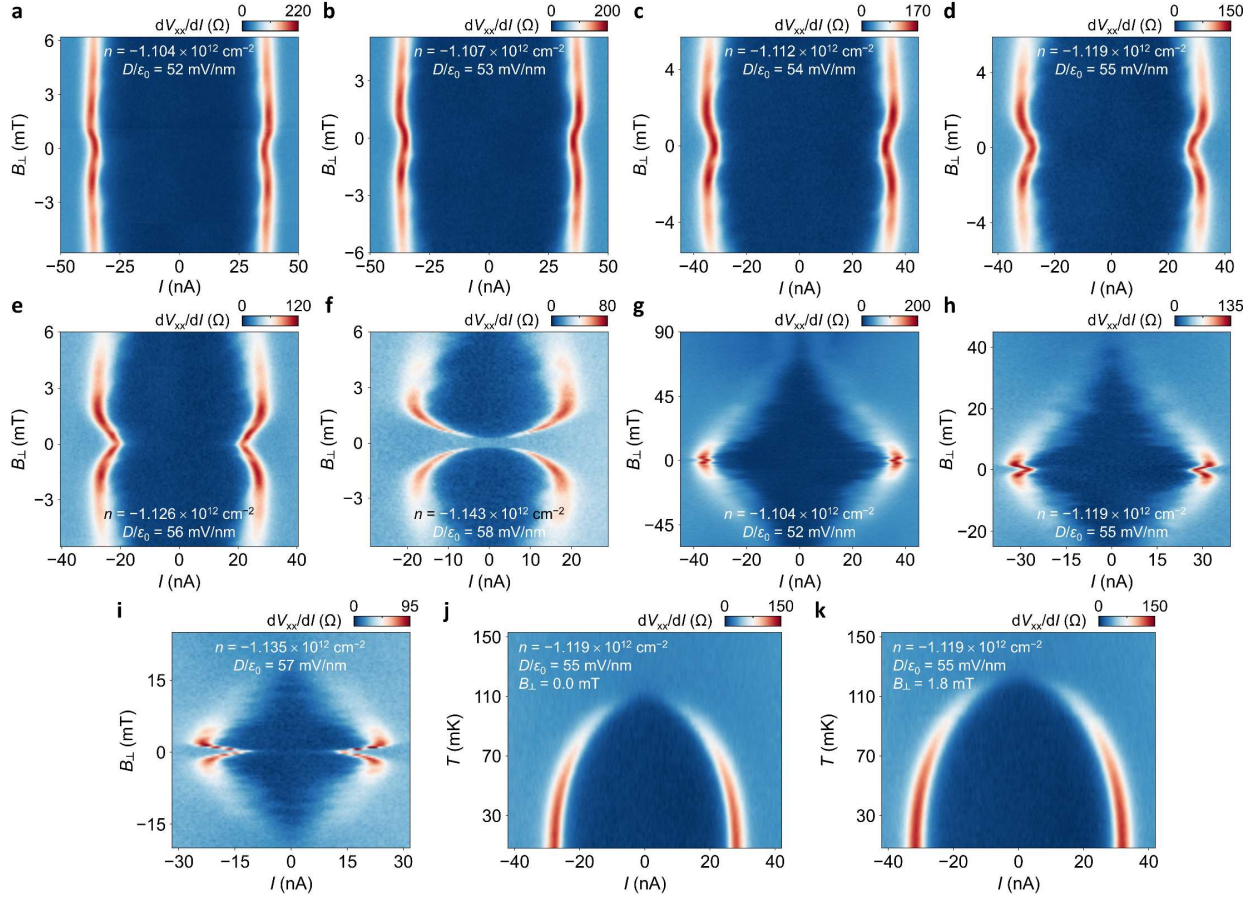

**Extended Data Figure 5. Enhancement of SC3 in R5G by $B_\perp$. a-f,** Differential resistance as a function of $I$ and $B_\perp$ at different $(n, D)$ points. The enhancement of SC3 by $B_\perp$ is observed globally across its phase space. **g-i,** Differential resistance as a function of $I$ and $B_\perp$ at different $(n, D)$ points for a wider range of the magnetic field. Fraunhofer patterns arising from quantum interference are observed in **i** which is adjacent to the phase boundary of SC3, and they become fainter as $D/\varepsilon_0$ decreases. **j,k,** Differential resistance as a function of $I$ and $T$ at $n = -1.119 \times 10^{12}$ cm$^{-2}$ and $D/\varepsilon_0 = 55$ mV/nm under $B_\perp = 0$ and 1.8 mT. $T_{BKT}$ increases from 99 to 110 mK by applying $B_\perp$ of 1.8 mT, which demonstrates the increase in $T_{BKT}$ happens globally.

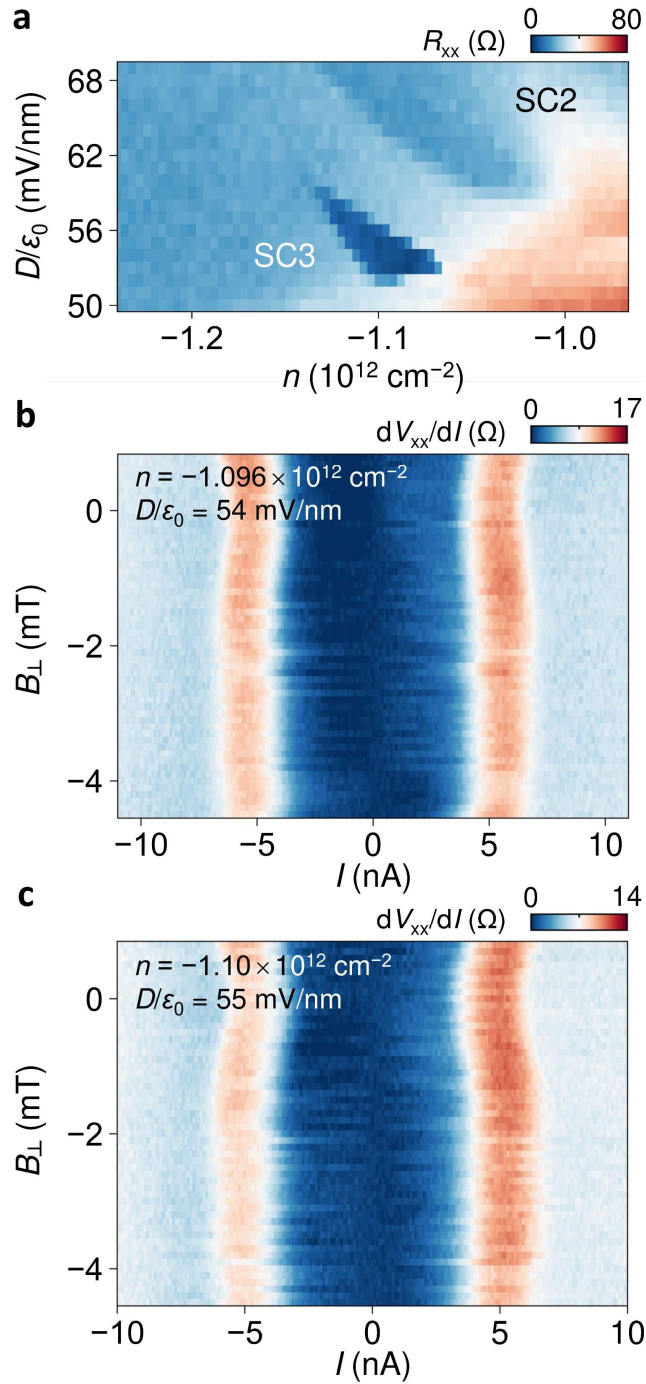

**Extended Data Figure 6. SC2 and SC3 in R5G from Device 2. a,** $R_{xx}$ map as a function of $n$ and $D/\varepsilon_0$ focusing on SC2 and SC3. **b,c,** Longitudinal differential resistance as a function of $I$ and $B_\perp$ at the two ($n$, $D$) points. The critical current becomes maximum at $B_\perp$ of around −1.6 mT, which is similar to a value from Device 1 in Fig. 3f.

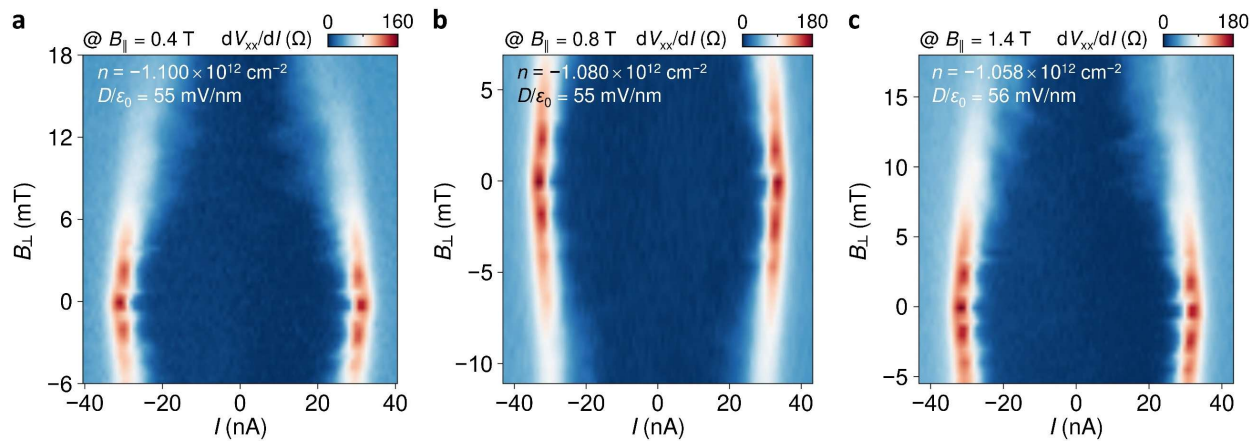

**Extended Data Figure 7. Absence of $B_\perp$-induced enhancement for SC3 in R5G under $B_\parallel$. a-c,** Differential resistance as a function of $I$ and $B_\perp$ for (**a**) $B_\parallel$ = 0.4 T, (**b**) $B_\parallel$ = 0.8 T and (**c**) $B_\parallel$ = 1.4 T.

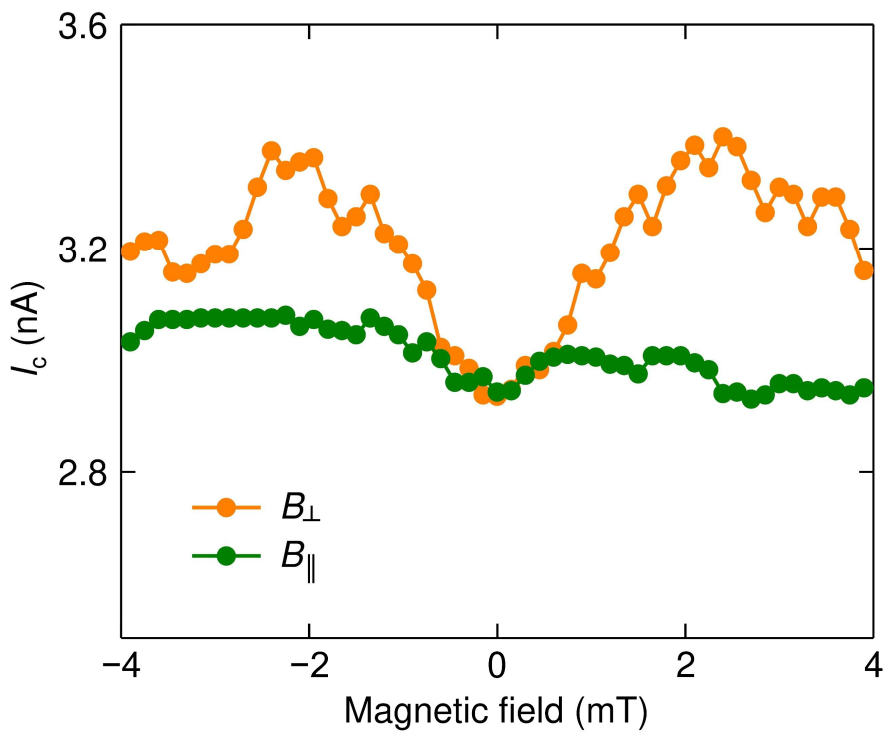

**Extended Data Figure 8. Dependence of the critical current of SC3 in R5G on $B_\perp$ and $B_\parallel$.** The critical current $I_c$ increases with showing peaks only when the out-of-plane field is applied.

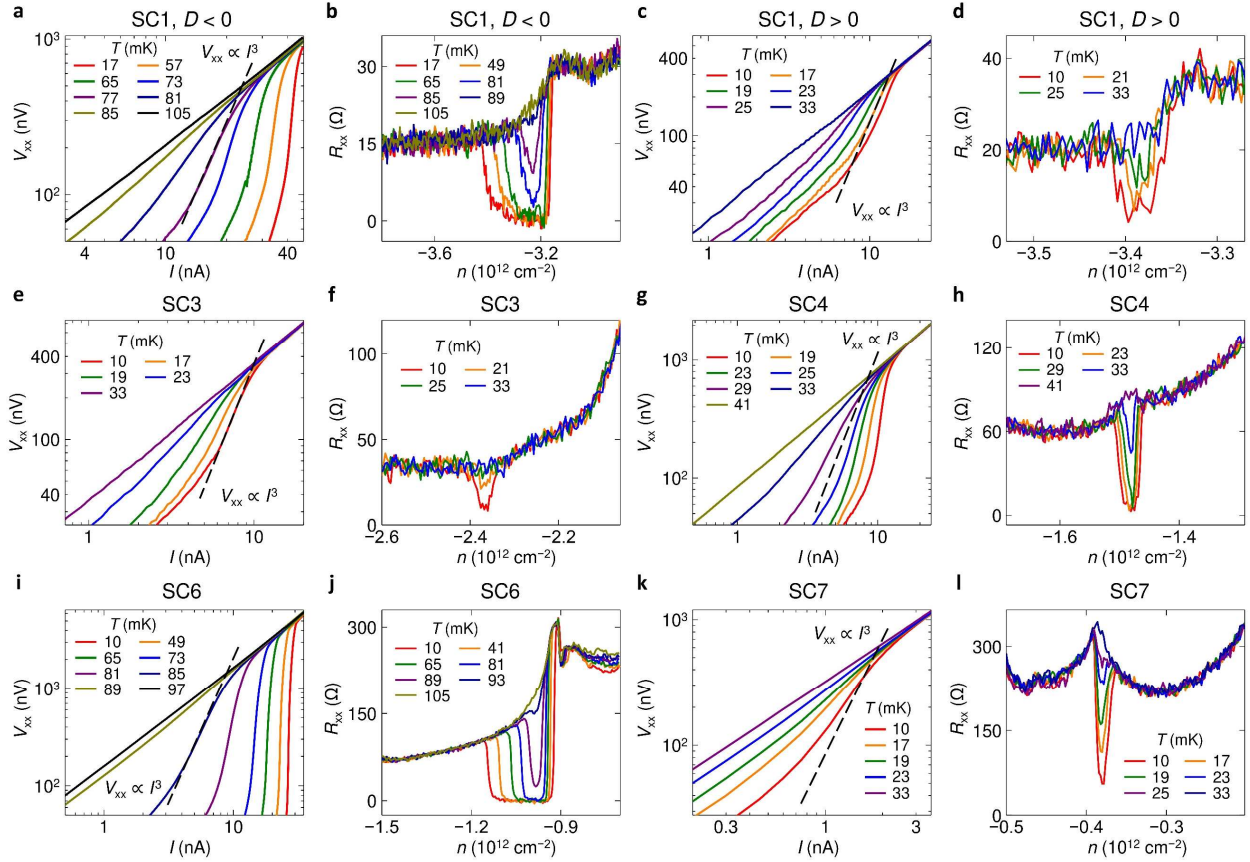

**Extended Data Figure 9. Temperature dependence of SC in R4G/WSe$_2$.** Differential resistance as a function of $I$ at varied temperatures for **a,** SC1 at negative $D$ ($n = -3.25 \times 10^{12}$ cm$^{-2}$, $D/\varepsilon_0 = -0.42$ V/nm), **c,** SC1 at positive $D$ ($n = -3.385 \times 10^{12}$ cm$^{-2}$, $D/\varepsilon_0 = 0.469$ V/nm), **e,** SC3 ($n = -2.356 \times 10^{12}$ cm$^{-2}$, $D/\varepsilon_0 = 0.18$ V/nm), **g,** SC4 ($n = -1.482 \times 10^{12}$ cm$^{-2}$, $D/\varepsilon_0 = 0.072$ V/nm), **i,** SC6 ($n = -1.00 \times 10^{12}$ cm$^{-2}$, $D/\varepsilon_0 = 0.026$ V/nm), and **k,** SC7 ($n = -0.38 \times 10^{12}$ cm$^{-2}$, $D/\varepsilon_0 = -0.24$ V/nm). The dashed lines indicate when $V_{xx}$ is proportional to $I^3$, which corresponds to the BKT transition. $R_{xx}$ as a function of $n$ at varied temperatures for **b,** SC1 at negative $D$ ($D/\varepsilon_0 = -0.4$ V/nm), **d,** SC1 at positive $D$ ($D/\varepsilon_0 = 0.469$ V/nm), **f,** SC3 ($D/\varepsilon_0 = 0.18$ V/nm), **h,** SC4 ($D/\varepsilon_0 = 0.072$ V/nm), **j,** SC6 ($D/\varepsilon_0 = 0.026$ V/nm), and **l,** SC7 ($D/\varepsilon_0 = -0.24$ V/nm).

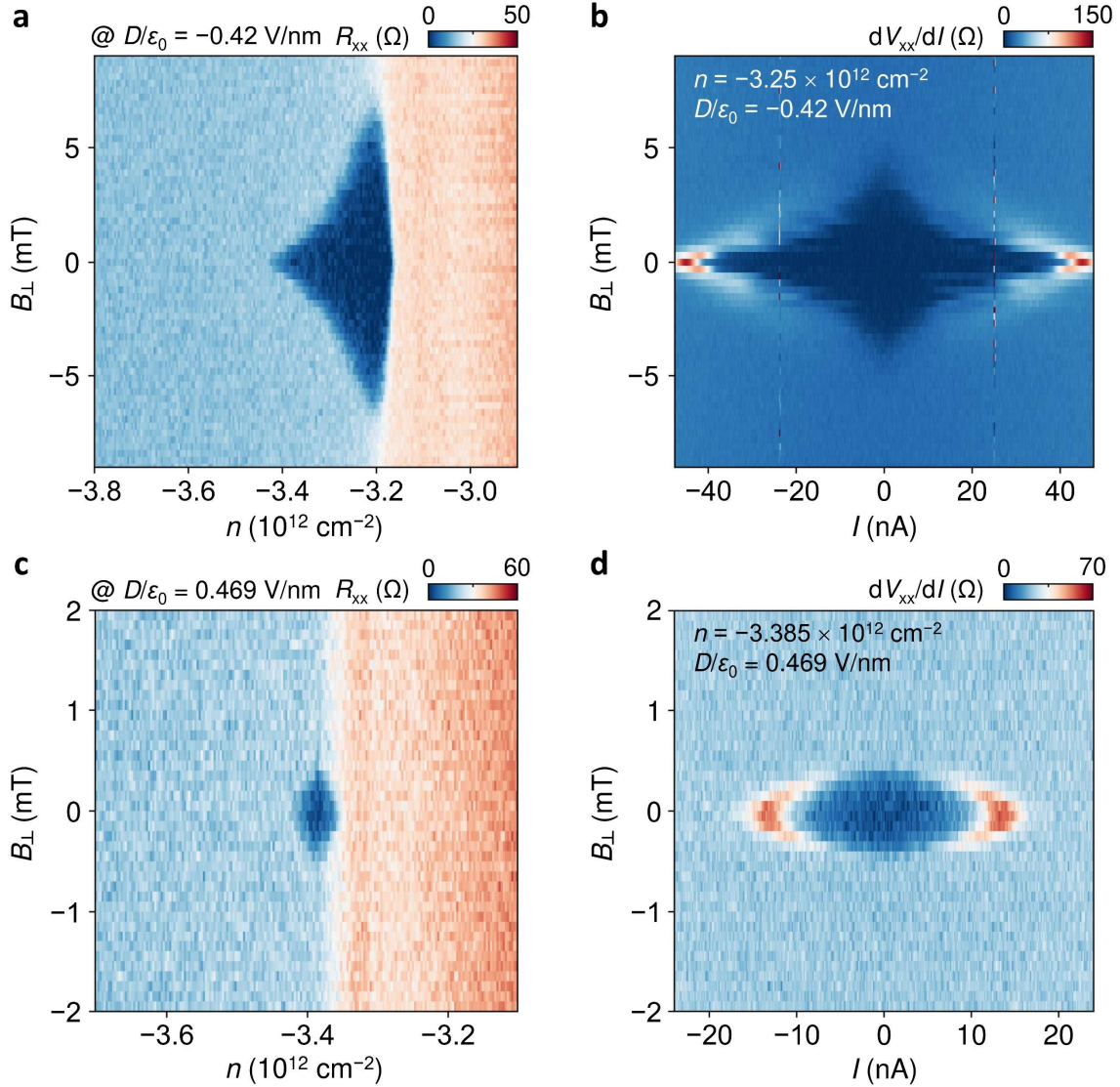

**Extended Data Figure 10. Comparison of SC1 in R4G/WSe₂ on the WSe₂-distant and WSe₂-proximal side. a,** $R_{xx}$ as a function of $n$ and $B_\perp$ for the SC1 on the WSe₂-distant side ($D/\varepsilon_0 = -0.42$ V/nm). **b,** Differential resistance as a function of $I$ and $B_\perp$ for the SC1 on the WSe₂-distant side ($n = -3.25 \times 10^{12}$ cm⁻², $D/\varepsilon_0 = -0.42$ V/nm). **c,** $R_{xx}$ as a function of $n$ and $B_\perp$ for the SC1 on the WSe₂-proximal side ($D/\varepsilon_0 = 0.469$ V/nm). **d,** Differential resistance as a function of $I$ and $B_\perp$ for the SC1 on the WSe₂-proximal side ($n = -3.385 \times 10^{12}$ cm⁻², $D/\varepsilon_0 = 0.469$ V/nm). These data show the range of SC1 along the $n$ axis, $I_c$, and $B_{c,\perp}$ all decrease when SC1 is placed to the WSe₂-proximal side.